\documentclass[aip, amsmath,amssymb,reprint]{revtex4-1}
\usepackage{graphicx,color,upgreek}
\usepackage{mathtools}
\usepackage{amsmath}
\usepackage{graphicx,color,natbib}
\usepackage{placeins}
\usepackage[dvipsnames]{xcolor}
\usepackage{lipsum}

\let\oldAA\AA
\renewcommand{\AA}{\text{\normalfont\oldAA}}

\newcommand{\Pecl}{\operatorname{\mathrm{P\kern-.08em e}}}

\begin{document}

\preprint{AIP/123-QED}

\title{The rheology of confined colloidal hard discs}

\author{Ian Williams}
\affiliation{Department of Physics, University of Surrey, Guildford GU2 7XH, UK}
\affiliation{H.H. Wills Physics Laboratory, Tyndall Avenue, Bristol, BS8 1TL, UK}
\affiliation{Centre for Nanoscience and Quantum Information, Tyndall Avenue, Bristol, BS8 1FD, UK}
\affiliation{School of Chemistry, University of Bristol, Cantock's Close, Bristol, BS8 1TS, UK}

\author{Erdal C. O\u{g}uz}
\affiliation{School of Mechanical Engineering, Tel Aviv University, Tel Aviv 6997801, Israel}
\affiliation{The Sackler Center for Computational Molecular and Materials Science, Tel Aviv University, Tel Aviv 6997801, Israel}
\affiliation{School of Chemistry, Tel Aviv University, Tel Aviv 6997801, Israel}
\affiliation{Key Laboratory of Soft Matter Physics, Institute of Physics, Chinese Academy of Sciences, Beijing 100190, China}

\author{Hartmut L\"{o}wen}
\affiliation{Institut f\"ur Theoretische Physik II, Heinrich-Heine-Universit\"at, D-40225 D\"usseldorf, Germany}

\author{Wilson C. K. Poon}
\affiliation{SUPA and School of Physics \& Astronomy, The University of Edinburgh, Peter Guthrie Tait Road, Edinburgh EH9 3FD, UK}

\author{C. Patrick Royall}
\email{paddy.royall@epsci.psl.eu}
\affiliation{Gulliver UMR CNRS 7083, ESPCI Paris, Universit\'{e} PSL, 75005 Paris, France}
\affiliation{H.H. Wills Physics Laboratory, Tyndall Avenue, Bristol, BS8 1TL, UK}
\affiliation{Centre for Nanoscience and Quantum Information, Tyndall Avenue, Bristol, BS8 1FD, UK}
\affiliation{School of Chemistry, University of Bristol, Cantock's Close, Bristol, BS8 1TS, UK}

\date{\today}

\begin{abstract}
\textbf{
Colloids may be treated as ``big atoms'' so that they are good models for atomic and molecular systems. Colloidal hard disks are therefore good models for 2d materials and although their phase behavior is well characterized, rheology has received relatively little attention. Here we exploit a novel, particle-resolved, experimental set-up and complementary computer simulations to measure the shear rheology of quasi-hard-disc colloids in extreme confinement. In particular, we confine quasi--2d hard discs in a circular ``corral'' comprised of 27 particles held in optical traps. Confinement and shear suppress hexagonal ordering that would occur in the bulk and create a layered fluid. We measure the rheology of our system by balancing drag and driving forces on each layer. Given the extreme confinement, it is remarkable that our system exhibits rheological behavior very similar to \emph{unconfined} 2d and 3d hard particle systems, characterized by a dynamic yield stress and shear-thinning of comparable magnitude. By quantifying particle motion perpendicular to shear, we show that particles become more tightly confined to their layers with no concomitant increase in density upon increasing shear rate. Shear thinning is therefore a consequence of a reduction in dissipation due to a weakening in interactions between layers as the shear rate increases. We reproduce our experiments with Brownian Dynamics simulations with Hydrodynamic Interactions (HI) included at the level of the Rotne--Prager tensor. That the inclusion of HI is necessary to reproduce our experiments is evidence of their importance in transmission of momentum through the system.}
\end{abstract}

\maketitle

\section{Introduction}

Confined fluids often exhibit modified flow behavior compared to the bulk due to coupling to the boundary. This may be static (\emph{i.e.} structural alterations due to energetic or entropic interactions) or dynamic (\emph{e.g.} the hydrodynamic influence of the wall). The shear viscosity of simple liquids increases by orders of magnitude in films less than $\sim 7$ molecules thick \cite{hu1991,demirel1996,klein1998}. This is a consequence of confinement-induced solidification near the boundary, driven by van der Waals interactions  \cite{horn1981,christenson1987,klein1995}. By contrast, the viscosity of water increases more modestly under similar conditions \cite{raviv2001,antognozzi2001} as confinement-induced solidification is suppressed by the hydrogen bond network.

On the mesoscopic scale, rheological measurements of two-dimensional soft materials typically focus on interfacially adsorbed components including nanoparticles \cite{maestro2015,feng2016}, proteins \cite{cicuta2003,ariola2006,allan2014,williams2018}, lipid \cite{kim2011,sachan2017,williams2019,kim2018}, surfactant \cite{zell2014} and polymer \cite{anseth2003,cappelli2016,feng2016,pepicelli2017} monolayers or asphaltene films \cite{fan2010,chang2018,chang2019}, 2d foams \cite{katgert2008,raufaste2009}, Hele-Shaw emulsions \cite{desmond2015} lipid bilayers \cite{hormel2014} dusty plasmas \cite{feng2010} and colloids \cite{cicuta2003,masschaele2011,buttinoni2015,deshmukh2015,buttinoni2017}. The latter can be tailored to exhibit interactions similar to simple models of atomic and molecular systems, and since they explore phase space in much the same way colloids provide a means to probe the same phenomena, for example the effect of confinement, and other external fields such as shear, upon phase behavior \cite{rice2009,ivlev,lowen2009}. Due to their mesoscopic size, colloids may be confined by walls that are smooth on the particle scale or by walls that are rough (like atomic and molecular systems).

Another class of materials at a somewhat larger lengthscale is (athermal) granular matter and here too individual particles may be studied, and even the force networks between them \cite{majmudar2005}. Under vibration, granular matter can mimic the phase behavior of thermal systems \cite{reis2006}. The rheology of granular matter has received considerable attention \cite{gdrmidi2004,foterre2008,daniels2005,dijksman2011,boyer2011,watanabe2011,mandal2017}.  Like suspensions of colloids, granular materials become very much more viscous at high packing fraction, however the limiting case of divergence viscosity is jamming in the case of grains, rather than the thermal glass transition (which occurs at a lower packing fraction) in the case of colloids \cite{ikeda2012}. At the microscopic level, the shear response of colloidal and granular systems is distinguished in that colloidal particles usually do not come into physical contact with each other or with the walls of the sample cell. Therefore, dissipation occurs through hydrodynamic coupling and Brownian motion rather than friction due to direct contacts as in granular matter. Very occasionally, under extreme shear rates for example, contact can occur, leading to very sudden shear thickening \cite{mari2015,morris2019}.

In colloidal systems at volume or area fractions where glassy dynamics are encountered, researchers report both increases and decreases in viscosity with respect to the bulk depending on the degree of confinement and the boundary details. Relaxation may be accelerated (decelerated) near smooth (rough) walls \cite{scheidler2003,nugent2007,sarangapani2008}. These effects are attributed to boundary-induced structural modification dependent on the shape, roughness, wetting characteristics or interaction potential of the wall (to name but a few possibilities) \cite{oguz2012,isa2007,huber2015}. As a general rule, the formation of well-defined particle layers is associated with faster relaxation or reduced viscosity.

The bulk rheology of a hard-sphere-like (or hard-disc-like) colloidal suspension depends on its volume (or area) fraction, $\phi$ \cite{meeker1997}, and the shear rate, $\dot{\gamma}$. At low shear rates, viscosity decreases with $\dot{\gamma}$ \cite{brader2010,cheng2011,pham2008,zackrisson2006}. For volume (area) fractions approaching the hard sphere (disc) glass transition, shear thinning occurs after the applied stress exceeds a yield stress on the scale of $k_\mathrm{B} T / a^3$, where $k_\mathrm{B} T$ is the thermal energy and $a$ is the particle radius \cite{bonn2017,zackrisson2006,pham2008,henrich2009,vandervaart2013,legrand2008}. Upon increasing $\dot{\gamma}$, hard sphere suspensions exhibit a Newtonian plateau followed by shear thickening at very high shear rates \cite{cheng2011,bender1996,dhaene1993,brader2010}. Shear thinning in these systems is attributed to stratification of particles decreasing resistance to flow, while shear thickening at high $\dot{\gamma}$ is a consequence of frictional particle surfaces \cite{guy2015}.

\begin{figure*}[htb]
\begin{center}
\centerline{\includegraphics[width=170mm]{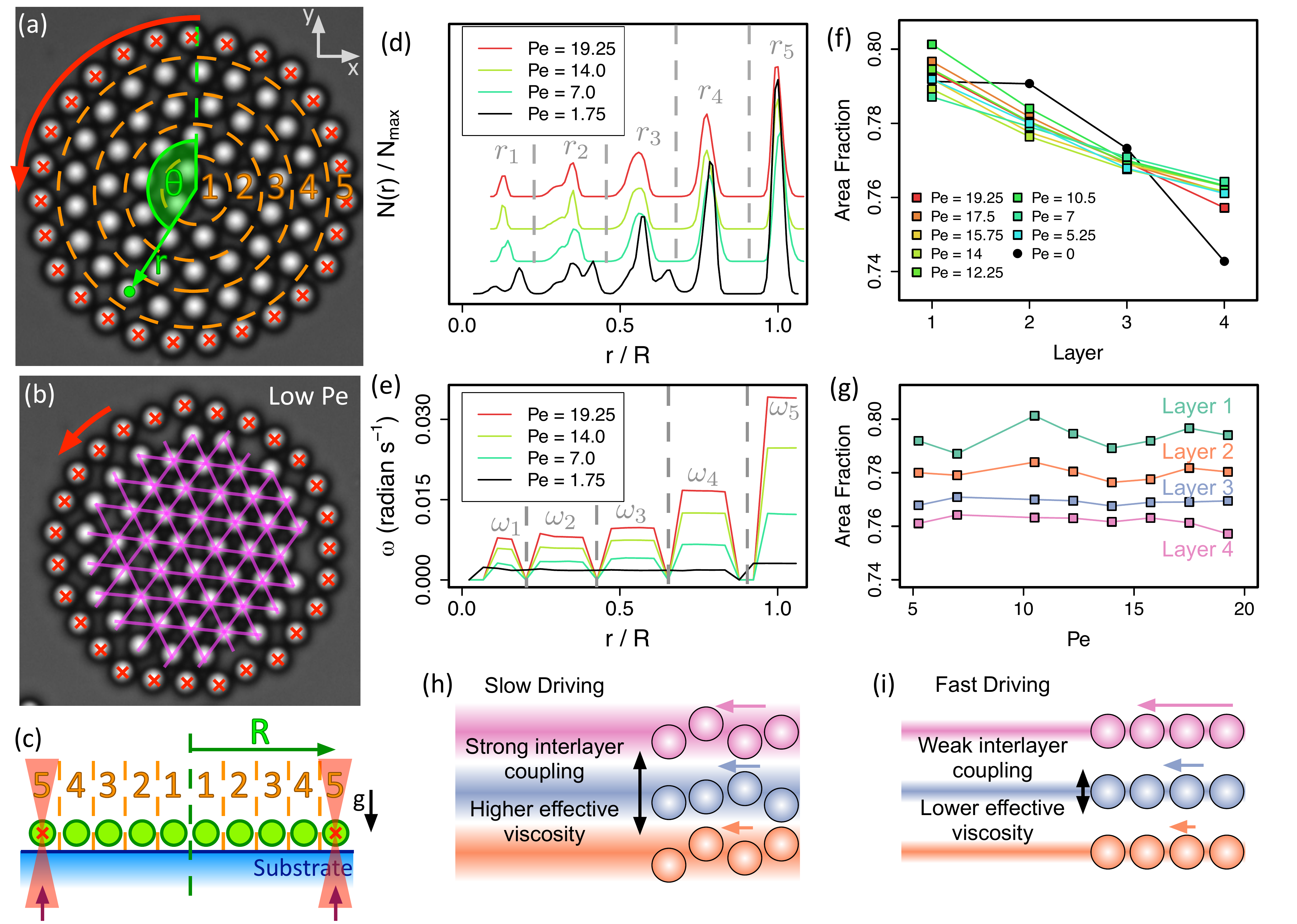}}
\caption{\label{figSchematic} 
(a-b) Annotated micrographs and (c) side-view schematic. 
(a) Shear-melted, layered system for $\Pecl \ge 5.25$. (b) Hexagonal structure at $\Pecl = 1.75$. Red crosses label optically trapped particles which are translated along the circular path indicated by the red arrow. Polar co-ordinates, $(r,\theta)$ are defined from the centre. $R$ is the boundary radius. Orange dashed lines in (a) and (c) demarcate particle layers, numbered 1 to 5 from the centre. Pink lines in (b) indicate hexagonal structure. 
(d) Histograms of radial particle location in experiments at $\Pecl$ indicated in legend. $y$-axis scale is arbitrary and data are offset in $y$ for clarity. 
(e) Angular velocity profiles corresponding to experiments in (d). Vertical dashed lines demarcate layers and zeros represent the inter-layer boundaries. 
(f) Area fraction as a function of layer number measured from average Voronoi cell areas in experiment. Lines are colored according to driving $\Pecl$. Black data represent the unsheared system where differing local structure leads to a change the distribution of local area fraction across the layers [see (d)].
(g) Area fraction from Voronoi cell area as a function of driving $\Pecl$ for each layer, measured in experiment. 
(h-i) Illustrated interpretation of shear thinning. Coupling between layers depends on the range of radial motion within adjacent layers. In a slowly driven system, (h), particles explore a larger radial extent than in a quickly driven system, (i). For larger local shear rate, particles in adjacent layers are further apart, interlayer interactions are weaker and the effective viscosity is lower. This manifests as shear thinning.	
}
\end{center}
\end{figure*}

While the rheology of (3d) colloidal hard spheres has been extensively studied \cite{besseling2007,schall2007,koumakis2012} attention to their 2d analogue, hard discs has focussed on quiescent (unsheared) systems  \cite{brunner2002,collins2015,stoop2018,brunner2003,stratton2009,thorneywork2014,tamborini2015,williams2015,gray2015,williams2018jpcm}. Here, we focus on the rheology of confined colloidal hard discs. We present experiments on a quasi-hard-disc system of spherical colloids adjacent to a solid substrate confined by 27 identical particles optically trapped along a circular boundary and subjected to shear by boundary rotation. These are accompanied by computer simulation of a 2d system using Brownian Dynamics with hydrodynamic interactions (HI) included at the level of the Rotne--Prager tensor. Under quiescent conditions, our system has a bistable state between a hexagonal configuration that distorts the flexible walls [Fig. \ref{figSchematic} (b)] and a layered fluid where hexagonal ordering is suppressed [Fig. \ref{figSchematic} (a)]. Hexagonal configurations exhibit voids adjacent to the wall, enabling relaxation mechanisms that are absent in the bulk \cite{williams2015}. Under shear, hexagonal configurations rotate as a rigid body, slipping only at the wall interface, while layered fluid configurations slip between each layer. At higher shear rates, the system is shear melted and only the layered fluid is %structures are 
observed \cite{williams2016}. It has subsequently been shown that simulations of a similar system exhibit distinct regions of shear thinning and thickening as a function of shear rate \cite{ortizambriz2018,gerloff2020}. Here, we develop a layer-by-layer approach to determine local rheological properties directly from experiments and simulations via drag forces, and we characterise motion perpendicular to shear via mean squared displacements. Our approach opens a route to particle-level analysis of rheological properties in experimental systems.

The key finding of this work is that, in contrast with many molecular liquids \cite{hu1991,demirel1996,klein1998}, strong confinement has little effect on the shear rheology of colloidal quasi-hard-discs. Our measurements are qualitatively and quantitatively aligned with simulations of hard discs under steady shear \cite{henrich2009}. We measure a flow curve indicative of a yield stress, and shear thinning towards a Newtonian plateau. Motion perpendicular to shear is suppressed with no change in local density as shear rate is increased. We hypothesize that this suppression of perpendicular motion reduces the coupling between adjacent particle layers, reducing drag and dissipation, and consequently we measure a reduction in viscosity. This interpretation is illustrated in Fig. \ref{figSchematic} (h-i). Our simulations reveal that the mechanism of momentum transfer in this system is dominated by hydrodynamic interactions, and the excluded volume interactions between the particles play a rather minor role.

This paper is organised as follows. Section \ref{secMethods} describes the experimental and simulation procedures, and details of our stress calculation. Full details of the drag coefficient determination, simulations and interlayer hopping are provided in the Appendices. Section \ref{secStructureAngVel} presents key structural and dynamic measurements required for the measurements of local shear rate, stress and viscosity in Section \ref{secRheoAnalysis}. Section \ref{secMSDanalysis} quantifies the motion perpendicular to shear. Finally, Section \ref{secDiscussion} combines these measurements to develop our explanation of shear thinning. We conclude this work in section \ref{secConclusion}.

\section{Methods}
\label{secMethods}

\subsection{Experiment}

Figures \ref{figSchematic} (a) and (b) show micrographs and (c) shows a side-view schematic of the system. 
Polystyrene spheres of radius $a = 2.5 \; \mathrm{\upmu m}$ and polydispersity 2\% are suspended in a $3:1$ mass ratio mixture of deionised water and ethanol, loaded into a cell constructed using three glass coverslips and a microscope slide and sealed with epoxy.  Owing to their density mismatch with the solvent, the particles quickly sediment, forming a quasi-two-dimensional layer adjacent to the lower glass coverslip, which is treated with Gelest Glassclad 18 to prevent particle adhesion. The gravitational length is $l_g / a = 0.030 \pm 0.002$, resulting in negligible out-of-plane particle motion in the $z$ direction, while still being far enough from the substrate that any direct interactions with the substrate through contacts can be safely neglected. Interparticle interactions are of Yukawa form with a Debye length $\lambda_\mathrm{D} \approx 25 \; \mathrm{nm}$, which is sufficiently short that they may be considered quasi-hard-discs \cite{williams2013}. In the dilute limit, the time for a particle to diffuse its radius (the Brownian time) in the substrate-adjacent, quasi-two-dimensional layer is measured to be $\tau_\mathrm{B} \approx 70 \; \mathrm{s}$.

Computer-controlled holographic optical tweezers in an inverted microscope are used to manually gather $N=75$ particles, $%n_5 = 
27$ of which are optically trapped to form a circular boundary of radius $R$ which confines the remaining $n_{\mathrm{conf}}=48$. The spring constant of the optical traps maintaining this boundary is extracted from the trapped particle trajectories and is determined to be $k = 105(1) \; k_\mathrm{B}T a^{-2}$. The colloids sediment to the bottom of a sample cell, and thus there is a (relatively) large amount of solvent above. We presume that any local heating effect is swiftly dissipated to the surrounding solvent. When we tested the system, for example at different strengths of the laser tweezers, we saw no signs of any such heating.

Using the optical tweezers, the boundary particles are translated along a circular path through the periodic displacement of the optical tweezers array in discrete steps of length $a/4$. The frequency with which the array of traps is updated defines the boundary rotation speed, which we characterise using the boundary P\'{e}clet number, $\Pecl = \tau_\mathrm{B} / \tau_D$, where $\tau_D$ is the time taken to drive a boundary particle a distance $a$. We consider experiments with $\Pecl$ in the range $1.75 \le \Pecl \le 19.25$.

Following the initiation of boundary rotation, the system is left to complete $5$ full rotations before micrographs are acquired at a rate of $2$ frames per second for up to $3$ hours. 
No indication of anything other than a steady-state behavior was found. Particle trajectories are extracted from micrographs using standard algorithms \cite{crocker1996}. Circular polar co-ordinates, $r$ and $\theta$, are defined from the system centre.

\subsection{Simulation}

We perform 2d Brownian Dynamics simulations of strongly screened charged colloids interacting via a Yukawa pair potential under highly ionic conditions. 
\begin{equation}
V(r)=V_0\dfrac{\mathrm{e}^{-\kappa r}}{\kappa r} ,
\label{eqYuk}
\end{equation}
with $r$ denoting the inter-particle separation.
The inverse screening length $\kappa$ is chosen as $\kappa a = 14.85$ and the contact potential $V(r=2a) \approx V_0 k_BT$ where $V_0=0.85$, informed by the experimental parameters, ensuring quasi-hard-disc behavior \cite{williams2016}. In particular, we found empirically that $\kappa a = 14.85$ was sufficiently hard to closely reproduce key dynamical properties of the experimental system such as the way in which layers of particles slip past one another. Further details are available in Appendix \ref{secSimulationDetails} and Ref.  \cite{williams2016}. We neglect polydispersity. Boundary particles experience harmonic potentials mimicking optical traps and are translated along a circular path at a prescribed rate. Hydrodynamic effects are accounted for at the Rotne-Prager level \cite{gauger2009,hansen2011} in the presence of a planar substrate (Blake's solution \cite{blake1971}), giving the overdamped equation of motion for a particle trajectory $\textbf{r}_i$ in a time step $\delta t$
\begin{equation}
\mathbf{r}_i (t+\delta t) = \mathbf{r}_i(t) + \left( \sum_{j=1}^{N} \tensor{\mathbf{\mu}}_{ij} \mathbf{F}_j \right) \delta t + \delta \textbf{W}_i.
\label{eq:eq8}
\end{equation}
The mobility tensor $\tensor{\mathbf{\mu}}_{ij}$ comprises both the self-mobility and the entrainment of particle $i$ by the hydrodynamic flow field created by conservative forces $\mathbf{F}_j$ on particle $j$. This force stems from pair interactions and, for the boundary particles, the harmonic potentials. The random displacement, $\delta \textbf{W}_i$, is sampled from a Gaussian distribution with zero mean and variance $2D_0 \delta t$ (for each Cartesian component) fixed by the fluctuation-dissipation relation, where $D_0$ is the diffusion coefficient. Further details are provided in Appendix \ref{secSimulationDetails}. We have previously shown that these simulations faithfully reproduce experiments both qualitatively and quantitatively \cite{williams2016}.

\subsection{Stress Calculation}
\label{secStressCalculation}

Figure \ref{figSchematic}(a) shows that the particles organize into concentrated layers. This motivates our determination of the stress, by considering each layer, $i$. We presume that the angular velocity and radial position are constant within each layer, which, as we shall show in Section \ref{secResults} is reasonable for our purposes. We thus perform a series of force balances to obtain the stress across each layer, $\sigma_i$. Within each layer there are three forces which must sum to zero at steady state. There is some driving force (either optical forces or the transmitted force due to the motion of an external layer), $F_{\mathrm{ext}}$. This is balanced by some self-drag force representing dissipation within the layer, $F_{\mathrm{self}}$, and some additional force that drives the next internal layer, $F_{\mathrm{int}}$. In the case of layer $1$ (the central layer), there is only self-drag.

The dilute limit single particle drag coefficient near the substrate, $\zeta_\mathrm{emp} = 1.9 \times 10^{-7} \; \mathrm{kg \, s^{-1}}$, is extracted from measurements of Brownian motion in the dilute limit, without any optical traps. However, multiple particles moving along a circular path of radius $r$ near a substrate each experience reduced hydrodynamic drag due to the presence of the other particles \cite{ladavac2005}. This is the \emph{drafting effect}. This drag reduction depends on the number of particles and $r$, which suggests that particles in different layers experience different drag coefficients. Based on the discussion in Appendix \ref{secDeterminingDrag} we assume a constant drag coefficient throughout the system, $\zeta = 0.34 \zeta_\mathrm{emp}$, which is reduced compared to the dilute limit drag coefficient due to many-body hydrodynamic effects, but is independent of radius. This effect is discussed in greater detail and this approximation is justified in the Appendix \ref{secDeterminingDrag}. We therefore assume that the self-drag force has the same form for all particles and is proportional to velocity, $F_{\mathrm{self}} = \zeta r \omega$.

Consider layer $5$, the boundary, which consists of $n_5 = 27$ particles, each of which is located at radial position $r_5$ and subjected to an optical force $F_\mathrm{opt}$. All $n_5$ particles move with angular velocity $\omega_5$ and each experiences a self-drag proportional to its velocity. Balancing the forces on layer $5$ yields:
\begin{equation}
\label{eq:layer5forces}
n_5 F_\mathrm{opt} - n_5 \zeta r_5 \omega_5 - F_4 = 0,
\end{equation}
where $F_4$ is the as of yet unknown force transmitted inwards to drive the motion of all $n_4 = 21$ particles forming layer $4$. The force balance for layer 4 is then:
\begin{equation}
\label{eq:layer4forces}
F_4 - n_4 \zeta r_4 \omega_4 - F_3 = 0,
\end{equation}
where, once again, $F_3$ is the unknown force required to drive layer 3. Propagating the layer-by-layer force balance inwards to layer $1$ reveals the retrospectively trivial result that the optical driving forces are exactly balanced by the self-drags experienced by all of the particles:
\begin{equation}
\label{eq:sumofdrags}
n_5 F_\mathrm{opt} - \zeta \sum_{i=1}^5 n_i r_i \omega_i = 0.
\end{equation}
Replacing the optical forces by the sum of the drag forces gives the unknown force required to drive layer $i$, $F_i$, as the sum of the drag forces on layer $i$ and all layers that are internal to $i$:
\begin{equation}
\label{eq:generalforce}
F_i = \zeta \sum_{j=1}^i n_j r_j \omega_j.
\end{equation}

Assuming that the inter-layer forces act at circular contact lines between layers at radial locations intermediate between the two layer centres, we calculate the two-dimensional tangential stress on layer $i$:
\begin{equation}
\label{eqStressj}
\sigma_i = \frac{\zeta}{\pi (r_i + r_{i+1})} \sum_{j=1}^i n_j r_j \omega_j.
\end{equation}

\section{Results}
\label{secResults}

We begin by recapitulating the structural and dynamic features necessary for our rheological analysis. We subsequently implement the analysis described in section \ref{secStressCalculation} to measure the shear rheology combining particle-level structural and dynamic information to extract stresses from drag forces. Finally, we quantify particle motion in the radial direction, perpendicular to shear.

\subsection{Angular Velocity Profiles and Structure}
\label{secStructureAngVel}

Without shear, the system can adopt either layered fluid or locally hexagonal structures, shown in Fig. \ref{figSchematic} (a) and (b) respectively \cite{williams2015}. When sheared at $\Pecl \lesssim 3$, hexagonal structures can persist and rotate as a rigid body \cite{williams2016}. The hexagonal structure is characterised by multiple peaks in the radial density profile [black line in Fig. \ref{figSchematic} (d)] and a flat angular velocity profile [black line in Fig. \ref{figSchematic} (e)]. For $\Pecl \gtrsim 3$, the system is shear melted and only layered structures are observed [coloured data in Fig. \ref{figSchematic} (d)]. The location of the $i$th peak in the layered density profile is labelled $r_i$. Layer populations are fixed at $n_1 = 3$, $n_2 = 9$, $n_3 = 15$, $n_4 = 21$ and $n_5 = 27$. In these shear melted experiments, angular velocity decreases in a step-like manner from the boundary to the centre coloured data in Fig. \ref{figSchematic} (e)], representing slipping between adjacent layers. The average angular velocity within layer $i$ is denoted $\omega_i$.

The area fraction is estimated for layer $i$ as $\phi_i = \pi a^2 / \langle A_\mathrm{V} \rangle_i$ where $\langle A_\mathrm{V} \rangle_i$ is the average Voronoi cell area for particles in layer $i$ \cite{gray2015}. Figure \ref{figSchematic} (e) shows that $\phi$ varies from $\phi_1 \approx 0.8$ at the centre, to $\phi_4 \approx 0.76$ adjacent to the boundary. At these densities, bulk hard discs are crystalline \cite{bernard2011}. However, in shear-melted experiments, hexagonal ordering is inhibited by the curved boundary and the application of shear and our system is liquid-like \cite{williams2015,williams2016}.

Shearing the system modifies the area fraction profile compared to the unsheared system [Fig. \ref{figSchematic} (e)], enhancing $\phi_4$ and suppressing $\phi_3$ and $\phi_2$. However, once the system is shear melted, the area fraction profile is insensitive to $\Pecl$ in the range of $\Pecl$ investigated, as shown in Fig. \ref{figSchematic} (f).

\begin{figure*}[!htb]
\begin{center}
\centerline{\includegraphics[width=170mm]{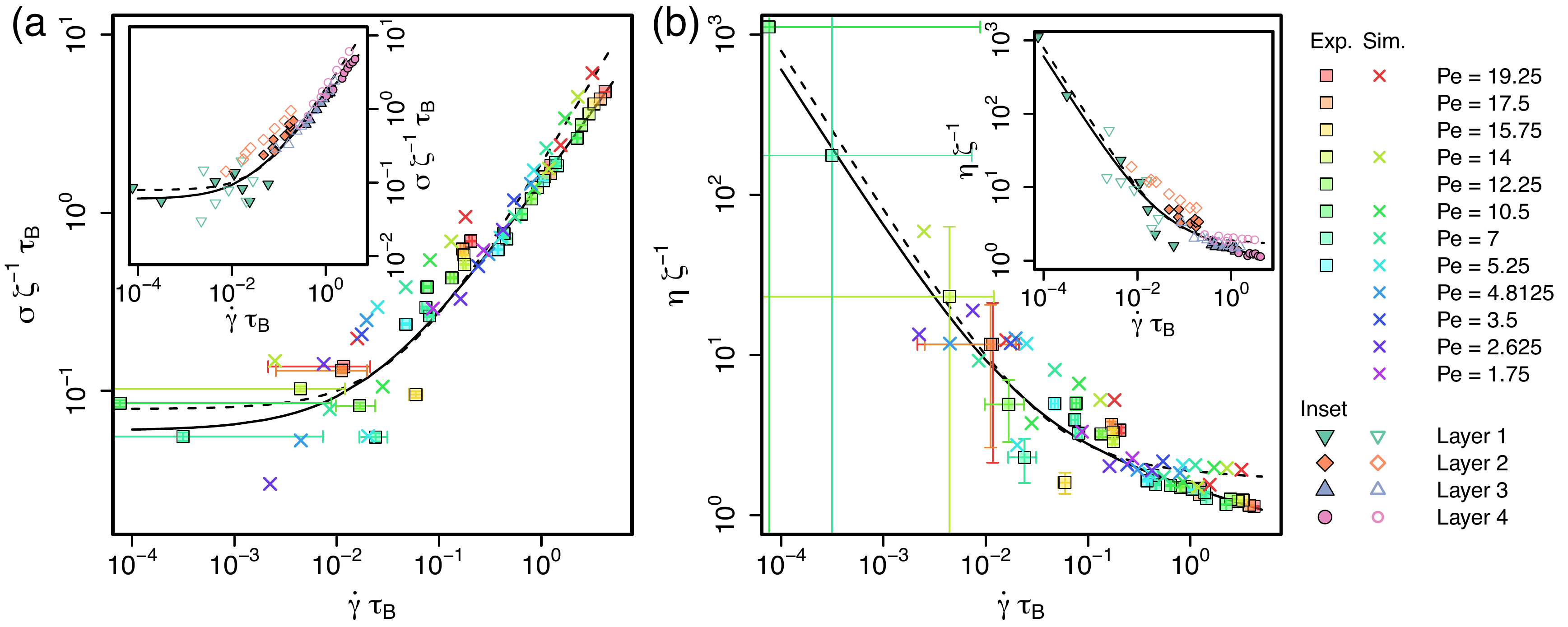}}
\caption{\label{figRheology} (a) Dimensionless stress and (b) viscosity as a function of dimensionless strain rate from experiments (squares) and simulations (crosses). Solid (dashed) line shows Herschel-Bulkley fit to experimental (simulated) data. Points are colored according to the driving $\Pecl$. Error bars are determined from angular velocity fluctuations in each layer. Insets show same data colored according to particle layer as indicated in the legend.}
\end{center}
\end{figure*}

\subsection{Shear Rheology}
\label{secRheoAnalysis}

To characterize the shear rheology, we follow the analysis in section \ref{secStressCalculation} and treat the shear-melted system as a series of coupled layers, following reference \cite{katgert2008}. Layers have fixed populations, $n_i$. We assume all particles are located at radial position $r_i$ corresponding to the peaks in the density profile [Fig. \ref{figSchematic} (d)] and move with angular velocity $\omega_i$ [Fig. \ref{figSchematic} (e)]. Viscosity is $\eta = \sigma/\dot{\gamma}$ where $\sigma$ is the stress and $\dot{\gamma}$ is the shear rate. The shear rate experienced by layer $i$ is obtained from the angular velocity profile as
\begin{equation}
\label{eqShearRateDiscrete}
\dot{\gamma}_i = r_i \frac{\Delta \omega_{i+1,i}}{\Delta r_{i+1,i}},
\end{equation}
where $\Delta \omega_{i+1,i} = \omega_{i+1} - \omega_i$ is the step down in $\omega$ between layer $i+1$ and layer $i$, and the radial separation between the layers $\Delta r_{i+1,i} = r_{i+1} - r_i$.

We determine the stress throughout the system as discussed in section \ref{secStressCalculation}. In particular, the corresponding stresses, $\sigma_i$, are obtained via a force balance on each layer, the key result is given in Eq. \ref{eqStressj}, that the force required to drive layer $i$ is the sum of the drag forces on layer $i$ and all layers internal to $i$. This force acts along a circular interlayer contact line, giving the stress on layer $i$, $\sigma_i$, from which we define an effective viscosity, $\eta_i = \sigma_i/\dot{\gamma}_i$, for each layer.

Figure \ref{figRheology} shows the results of this analysis for experiments (squares) and simulations (crosses). Stress is scaled by $\zeta^{-1} \, \tau_\mathrm{B}$, shear rate by $\tau_\mathrm{B}$ and viscosity by $\zeta^{-1}$ to facilitate the comparison. Points in the main panels are coloured according to the driving $\Pecl$, while the insets present the same data coloured by the layer number. Error bars are determined from angular velocity fluctuations in each layer determined from particle tracking. We see in Fig. \ref{figRheology} that the simulations do not reach such low shear rates low rates as the experiments. Now the only parameter that is directly set in both the experiments and the simulations is the rotation speed of the outer particle layer. Shear rates experienced by internal layers are determined by the physics of the system, ie the coupling between adjacent layers. So, the shear rates plotted are measurements, not directly controlled quantities.

The viscosity of hard particle suspensions depends on shear rate and volume/area fraction \cite{meeker1997,brader2010,cheng2011,pham2008,zackrisson2006,vandervaart2013,legrand2008,henrich2009}. The shear rate is largest in layer 4 and smallest in layer 1 and lower viscosity is measured for greater shear rate. However, Fig. \ref{figSchematic} (f) and (g) show that local area fraction is largest in layer 1 and decreases towards the boundary. Therefore, measuring larger viscosity nearer the centre of the system could be a consequence of increased density, and unrelated to the local shear rate. However, Fig. \ref{figSchematic} (g) shows that the area fraction of each layer is independent of $\Pecl$, and the inset to Fig. \ref{figRheology} (b) shows that the viscosity of a given layer \emph{does} decrease as the shear rate increases. We expect the spatial variation in area fraction to make some contribution to the viscosity, but the shear rate dependence is unambiguous.

The data in Fig. \ref{figRheology} are remarkably consistent with theoretical predictions and simulations of unconfined binary hard discs under steady shear \cite{henrich2009}. The flow curve suggests a dynamic yield stress \cite{fuchs2005,varnik2006,vandervaart2013}, beyond which, shear thinning is found. At high $\dot{\gamma}$, the response approaches a Newtonian region. The solid (dashed) black lines show Herschel-Bulkley fits to the experimental (simulated) data of the form $\sigma = \sigma_\mathrm{y} + k \dot{\gamma}^\nu$, where $\sigma_\mathrm{y}$ is the dynamic yield stress and $\nu$ is the high shear rate exponent, which is unity for Newtonian flow. The fits yield $\nu=0.82\pm0.03$ in experiment and $\nu=0.97\pm0.06$ in simulation, consistent with weakly shear thinning (experiment) and Newtonian (simulation) flow at the largest shear rates studied. We relate the yield stress to the change in structure upon the application of shear, from a hexagonal to layered fluid configuration. It is possible to fit the data in Fig. \ref{figRheology} with a power--law $\sigma = k' \dot{\gamma}^{\nu'}$ (rather than the Herschel--Bulkeley fit). While the error bars in the case of the data points corresponding to low shear rates are large enough that the this regime can be fitted with a power law, in fact the fit at high shear rate is much worse than the Herschel-Bulkeley fit shown. Further dependencies are of course possible, such as a power--law with a shear--rate dependent exponent. While this would be most interesting to explore in the future, the quality of our existing data means that it may be challenging to accurately discriminate between such more complex dependencies.

The dynamic yield stress is $\sim 0.1 k_\mathrm{B}T / a^2$. A dynamic yield stress is the stress required to maintain flow, and is smaller than the static yield stress, which must be exceeded during flow start-up. Although we focus on yielded systems, at very low $\Pecl$ the system is unyielded and rotates as a rigid body \cite{williams2016}. Rigid-body rotation requires the local stress to be less than the static yield stress. Therefore, the stress measured in unyielded systems is a lower limit estimate of the static yield stress. In experiment, the largest stress for which the system does not yield, is $\sim 1.9 \, k_\mathrm{B}T/a^2$, which is comparable to yield stresses at the scale $k_\mathrm{B} T / a^3$ in hard spheres \cite{zackrisson2006,pham2008,vandervaart2013,legrand2008}.

\begin{figure*}[htb]
\begin{center}
\centerline{\includegraphics[width=170mm]{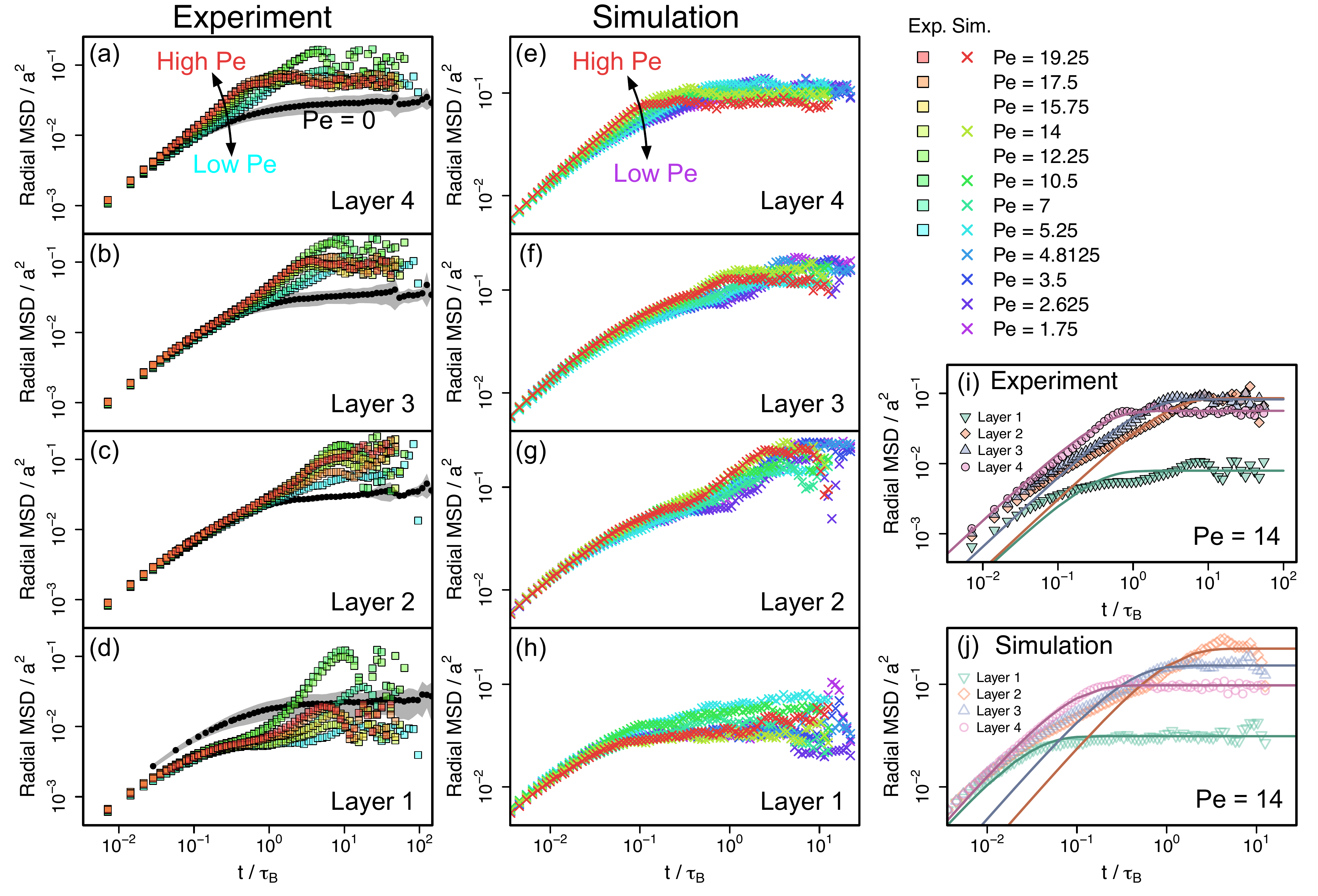}}
\caption{\label{figRadialMSDs} Radial mean squared displacements. All experimental data in (a) layer 4, (b) layer 3, (c) layer 2, and (d) layer 1. Black points and grey shaded regions represent average and standard deviation of behaviour in 5 unsheared experiments. All simulation data in (e) layer 4, (f) layer 3, (g) layer 2, and (h) layer 1. (i) All layers in a single experiment driven at $\Pecl = 14$.  (j) All layers in a single free centre simulation driven at $\Pecl = 14$. Points in (a-h) coloured according to driving $\Pecl$ and in (i) and (j) according to layer number as indicated in the legends. Lines in (i) and (j) show fits to Eq. \ref{eqKVMSD} as
 described in the text.}
\end{center}
\end{figure*}

The crossover from shear thinning to Newtonian flow occurs in the range $\dot{\gamma} \tau_B \sim 10^{-1}$ to $10^{0}$, which is coincident with this crossover in simulations and theory of hard discs \cite{henrich2009}, hard spheres \cite{cheng2011}, and charged colloids \cite{laun1984,kawasaki2014}. That the rheology under such strong confinement bears even a qualitative resemblance to bulk behaviour is a remarkable finding. Many molecular liquids and complex fluids exhibit increases in viscosity of many orders of magnitude when confined to a few particle layers \cite{hu1991,demirel1996,klein1998,scheidler2003,nugent2007,sarangapani2008}, but this is not the case for colloidal hard discs, which reproduce their predicted bulk rheology when confined to a system only $8$ particles across. We emphasize that the yield stress is likely related to the shear--induced melting of hexagonal order in this system. This ordering we have investigated previously \cite{williams2013,williams2016}.

\subsection{Motion Perpendicular to Shear}
\label{secMSDanalysis}

At the particle population of interest, and on the timescale of the experiment, particles are confined to layers (Appendix \ref{secInterlayer}). However, within the layers, they exhibit positional fluctuations in the radial direction as indicated by the width of the peaks in Fig. \ref{figSchematic} (d). The mean squared displacement (MSD) in the radial co-ordinate is shown in Fig. \ref{figRadialMSDs} for all experiments (a-d) and simulations (e-h). Black points in the experiment panels show the radial MSD for $\Pecl = 0$, averaged over 5 experiments. These MSDs grow with time up to a plateau which represents the confinement within layers. In the case of the experiments, the increase to a (somewhat noisy) plateau is continuous, in %the 
some simulation data, notably for layer 2, there is some evidence of two timescales in the radial MSD [Fig. \ref{figRadialMSDs}(g)]. Given that this two timescale behavior is only evidenced in layer 2 of the simulations, and not at all in the experiments, we focus on a single timescale, $\tau_\mathrm{rad}$ as determined below in Eq. \ref{eqKVMSD}.

Within each layer, the radial MSD approaches its plateau more quickly as $\Pecl$ is increased, indicating a coupling between radial and tangential motion. The shear rate is largest in layer 4 and decreases towards the system centre. The long-time plateau is reached most quickly in layer 4, and progressively more slowly in layers 3 and 2. Therefore, larger $\dot{\gamma}$ causes faster radial motion. Additionally, in all but the central layer, shearing increases the plateau above that measured in the unsheared system (black data) indicating that the amplitude of radial motion is increased under shear.

\begin{figure*}[htb]
\begin{center}
\centerline{\includegraphics[width=140mm]{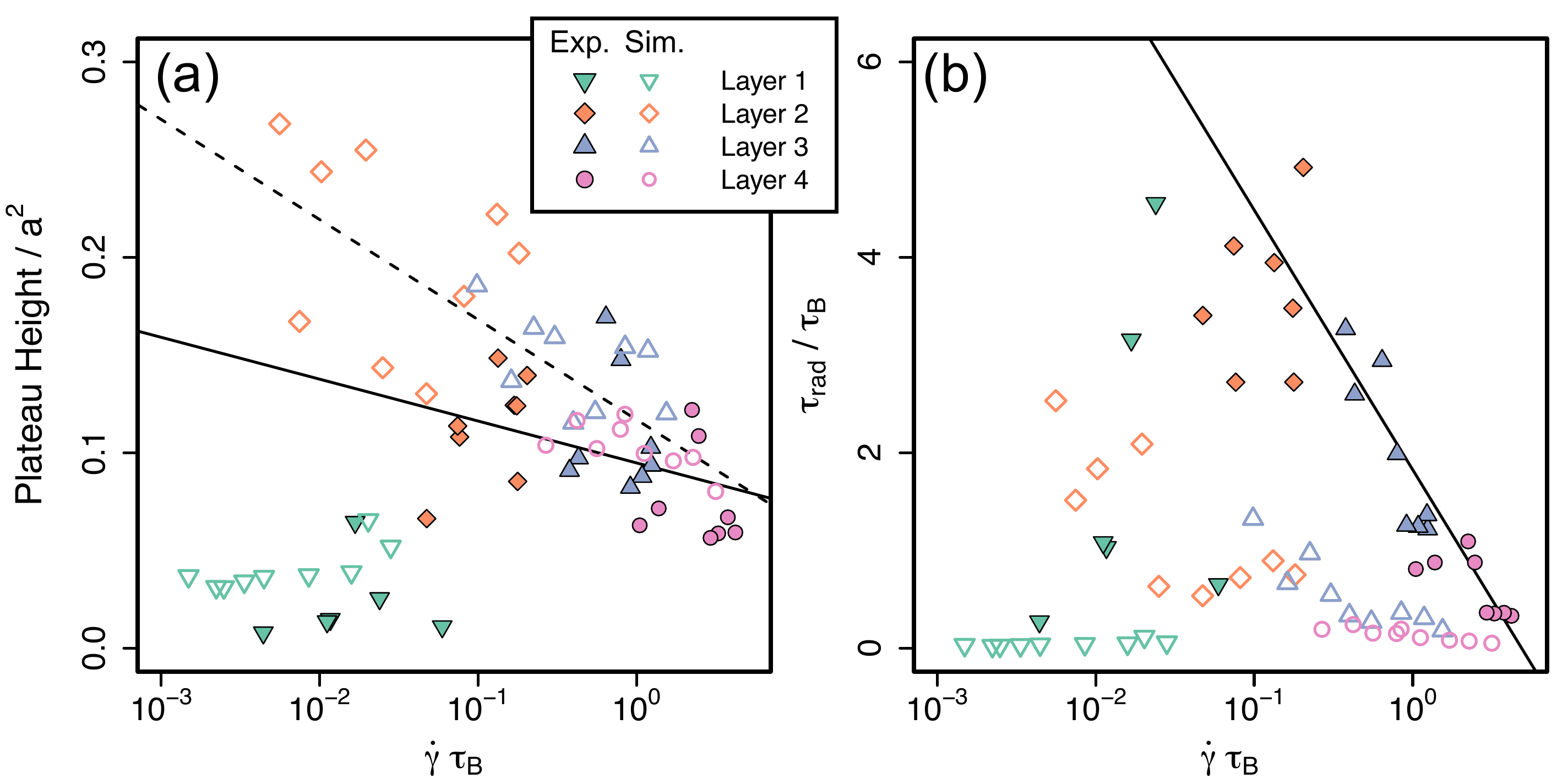}}
\caption{\label{figRMSDFits} Parameters extracted from fits to radial MSDs as a function of shear rate. (a) Plateau height $A$ (Eq. \ref{eqKVMSD}). (b) Timescale $\tau_\mathrm{rad}$ (Eq. \ref{eqKVMSD}). Solid (dashed) lines guide the eye to trends in experimental (simulated) data. Points are colored according to layer number indicated in the legend.}
\end{center}
\end{figure*}

Radial MSDs are fit with a function of the form  
\begin{equation}
\label{eqKVMSD}
\langle \Delta r^2(t) \rangle = A\left( 1 - e^{-t/\tau_\mathrm{rad}}\right),
\end{equation}
which captures their growth with $A$ the plateau height. Example fits for all four layers in an experiment and a simulation at $\Pecl=14$ are shown in Fig. \ref{figRadialMSDs} (i) and (j). We note that not all of the fits in (i) and (j) are of high quality. While we did find an improved fitting with the use of two timescales (i.e. two independent contributions to the right hand side of Eq. \ref{eqKVMSD}), in fact the small timescale proved to be very scattered, and its physical significance was unclear. A more sophisticated treatment than we have performed here with Eq.  \ref{eqKVMSD} would be interesting to explore in the future.

Figure \ref{figRMSDFits} shows the radial MSD plateau height $A$ and timescale $\tau_\mathrm{rad}$ as a function of the local shear rate for all experiments and simulations. The fit to the plateau height [Fig. \ref{figRMSDFits} (a)] reveals a downward trend with increasing $\dot{\gamma}$ in both experiment and simulation for all layers except layer 1. Figures \ref{figRadialMSDs} (a-d) show that shear enhances radial motion compared to the unsheared case (with the exception of layer 1). But, fitting reveals that this enhancement is greatest for lower shear rates. Particles are maximally radially mobile at the onset of shear melting to a layered fluid, and become increasingly confined in their layers at higher $\dot{\gamma}$.

The timescale $\tau$ shows a similar dependence on $\dot{\gamma}$ [Fig. \ref{figRMSDFits} (b)], decreasing as $\dot{\gamma}$ increases, indicating a faster approach to the long-time plateau. When subjected to faster shear, particles more quickly explore their full range of motion perpendicular to shear. This is very clearly evident for layers 2 to 4 in experiments. Once again, layer 1 does not follow the trend evident in layers 2 to 4.

\section{Discussion}
\label{secDiscussion}

The rheology of our system is similar to that of bulk colloidal hard discs, with evidence for a yield stress and shear thinning \cite{henrich2009}. Our particle-resolved approach allows us to address the origins of this yield stress and shear thinning. The former we believe is related to the breakdown of hexagonal configurations found in the absence of shear. The latter seems to be similar to the 3d case of shear-induced stratification \cite{cheng2011}. We have shown that motion parallel and perpendicular to shear are coupled. At larger local shear rate, particles more quickly explore their layer radially. Simultaneously, the extent of radial motion within the layer is reduced and they are more tightly confined. This is concurrent with shear thinning.

In colloidal hard sphere (and disc) systems, shear thinning is attributed to increased stratification parallel to shear as shear rate is increased \cite{wagner2009,kawasaki2014}. When particles are organised into layers, interactions between them become weaker and consequently the resistance to flow is reduced. Concentric particle layers are enforced by the boundary of our system, and therefore the flow resistance is inherently lower than in a bulk suspension at comparable density. However, Fig. \ref{figRheology} shows that shear thinning \emph{is} observed. We explain this fact using the data of Section \ref{secMSDanalysis}. Radial motion is suppressed as shear rate is increased, but Fig. \ref{figSchematic} (f) shows that local area fraction is independent of $\Pecl$. Faster shearing does not increase the packing density of particles, yet they exhibit more tightly confined dynamics. Radial motion brings particles to interact with particles in adjacent layers, which dissipates energy and resists flow. Suppressed radial motion with no concomitant increase in area fraction means that particles in adjacent layers are, on average, further apart and their interactions with one another are weakened. This results in a reduction in drag between layers, and therefore flow resistance, or viscosity. This interpretation is illustrated in Fig. \ref{figSchematic} (h) and (i). Thus, the radial dynamics suggest the rheology --- at larger shear rates, particles are more tightly confined to their layers at constant density. This reduces interlayer drag and leads to shear thinning.

What is the nature of the interactions that dominate the rheology of our system? Our simulations include hydrodynamic interactions at the Rotne--Prager level for the particle--particle interactions and also the Blake tensor for the coupling to the substrate. We found it necessary to include HI at this level to reproduce the behavior of the experiments. Pure Brownian Dynamics (without HI between the particles) leads to weak momentum transfer through the system, due to a far higher level of slip between the layers than is encountered in the experiments. Therefore we conclude that momentum transfer is dominated by HI and that steric effects due to %from 
the excluded volume of the particles play a secondary role. Thus, compared to molecular systems, where van der Waals interactions can lead to solidification near the boundary, here instead there are two important differences: (\emph{i}) there is no equivalent of the long--ranged van der Waals interactions in our hard discs, with no mechanism for solidification. (\emph{ii}) The momentum transfer is dominated by solvent--mediated hydrodynamic coupling, which is absent in molecular systems. The importance of hydrodynamic coupling between the particles suggests that similar behavior might be encountered in \emph{wet} granular matter \cite{mari2015,morris2019}. However in our system, the lack of direct particle--particle contacts (and increased ordering) enables a purely shear--thinning regime.

\section{Conclusion}
\label{secConclusion}

We have investigated the rheology of colloidal hard discs confined in a \emph{layered fluid} configuration under shear and hexagonal configuration with very weak or no shear in experiment and simulation. Using a particle-level analysis, we infer the inter-layer forces from drag forces and the driving force exerted by optical tweezers. Since this system is dissipative and in steady state, balancing the forces on each layer allows the measurement of the local viscosity. This is the first experimental measurement of the viscosity of a hard-disc-like colloidal system under steady shear.

The flow curve is indicative of a dynamic yield stress and shows that the confined hard disc system exhibits shear thinning at low shear rates and approximately Newtonian behavior for $\dot{\gamma} \tau_\mathrm{B} \gtrsim 0.1$. The dynamic yield stress is $\sim 0.1 \, k_\mathrm{B}T/a^2$. We find evidence for a static yield stress with a lower bound of $1.9 \, k_\mathrm{B}T/a^2$. This is remarkably similar to sheared, unconfined, bidisperse hard discs and bulk hard spheres. This is by no means an anticipated result as strongly confined systems regularly exhibit very different rheological responses to their bulk counterparts \cite{hu1991,klein1998,nugent2007,sarangapani2008}. In our system, there is a change in structure, in that the system undergoes shear melting, which we relate to the yield stress. In the future, it would be intriguing to explore the response of this system to oscillatory shear and to investigate any yield stress in more detail.

Shear thinning in colloidal hard particle systems is due to shear-induced layering progressively reducing off-axis interparticle collisions as shear rate is increased, reducing the viscosity. By measuring the particle motion perpendicular to the direction of shear, we show that this is also the case in our system. At higher shear rate, radial motion is suppressed without a change in local density and therefore interlayer particle collisions are reduced, reducing the coupling and dissipation between layers, and the therefore the viscosity.

Colloidal hard discs and spheres under extreme confinement behave remarkably similarly to their bulk counterparts in both 2 and 3 dimensions and are qualitatively different to systems dominated by van der Waals interactions, for which viscosity increases massively on increasing confinement \cite{hu1991,demirel1996,klein1998}. We attribute this to an absence of long--ranged vdW interactions in our system. Furthermore, we infer from our computer simulations that hydrodynamic coupling between the particles is the dominant mechanism of momentum transfer with excluded volume interactions playing a secondary role. This latter observation suggests that similar behavior might be observed in \emph{wet} granular matter, in the case that interactions due to contacts between particles are not dominant \cite{mari2015,morris2019}.

We have considered a particular geometry here, where a population of quasi-hard discs are effectively ``corralled'' by 27 tweezer particles arranged in a circle. Of course other geometries are possible. In the case of planar shear. we expect that behavior we observe would also be found, as one would expect particles to form layers parallel to the confinement, as is the case here. Depending on the particle spacing one might expect coupling between the packing of free particles and of wall particles.

In the future, it would be attractive to carry out a more complete inclusion of the HI than we have done here, for example with Lattice-Boltzmann dynamics or Stochastic Rotation dynamics. In particular, it would be useful to enquire whether such a description would exhibit the shear--thinning behavior seen in the experiments, and furthermore to explicitly probe lubrication phenomena neglected in the simulations we have performed here. It would also be interesting to develop a better description than the fit we have used to describe the radial MSDs in Eq. \ref{eqKVMSD}.

\section*{Acknowledgements}
The authors are grateful for enlightening discussions with Paul Bartlett, Wuge Briscoe, Olivier Dauchot, Jens Eggers, Yael Roichman, Thomas Speck and Todd Squires. HL was supported by the German Research Foundation (DFG) within the project LO 418/20-2. IW and CPR acknowledge support from the European Research Council (ERC Consolidator Grant NANOPRS, project number 617266).

\appendix{}

\section{Simulation Details}
\label{secSimulationDetails}

We perform Brownian dynamics simulations of $N$ particles interacting via a Yukawa pair potential (Eq. \ref{eqYuk}).
%\begin{equation}
%V(d)=V_0\dfrac{\mathrm{e}^{-\kappa d}}{\kappa d} ,
%\label{eq:eq1}
%\end{equation}
%with $d$ denoting the inter-particle separation, $\kappa$ the inverse Debye screening length, and $V_0$ the magnitude of the potential energy. 
Additionally, each of the $27$ particles in the outermost boundary layer is exposed to a harmonic potential mimicking the optical traps employed in experiment, given by
\begin{equation}
V_t (|\mathbf{r}_i-\mathbf{r}_{i,0}|)= \dfrac{k}{2} |\mathbf{r}_i-\mathbf{r}_{i,0}|^2 ,
\end{equation}
where $\mathbf{r}_i$ is the position of $i$th particle and $\textbf{r}_{i,0}$ the center of its potential well, with $k$ denoting the trap strength. At each time step $\delta t$, the locations of the $27$ harmonic potential minima, $\mathbf{r}_{i,0}$, are translated a predetermined arc length, $l$, along the boundary, resulting in a rotation velocity $l/\delta t$. The velocity, and thus the P{\'e}clet number, is controlled by altering this arc length.

The hydrodynamic interactions between colloids of radius $a$ are modeled on the the Rotne-Prager level \cite{gauger2009,hansen2011}. To account for the  hydrodynamic effect of the planar substrate that is present in experiments we first consider Blake's solution $\tensor{\mathbf{G}}^B$ \cite{blake1971}, which uses the method of images to obtain the Green's function of the Stokes equation satisfying the no-slip boundary condition at $z=0$. Furthermore, the hydrodynamic entrainment effect of the motion of particle $j$ at $\mathbf{r}_j$ on another particle $i$ at $\mathbf{r}_i$ is approximated by a multipole expansion \cite{kim2006,hansen2011} to the second order in $a$ leading to the Rotne-Prager level of the Blake tensor
\begin{eqnarray}
\tensor{\mathbf{G}}^{\mathrm{RPB}} (\mathbf{r}_i, \mathbf{r}_j)  & \equiv &  \left( 1 + \dfrac{a^2}{6} \nabla^2_{\mathbf{r}_i} \right. \nonumber + \left. \dfrac{a^2}{6} \nabla^2_{\mathbf{r}_j} \right) \tensor{\mathbf{G}}^\mathrm{B} (\mathbf{r}_i, \mathbf{r}_j) \nonumber \\
&=& \tensor{\mathbf{G}}^{\mathrm{RP}} (\mathbf{r}_{ij}) - \tensor{\mathbf{G}}^{\mathrm{RP}} (\mathbf{R}_{ij}) \nonumber  + \tensor{\Delta\mathbf{G}} (\mathbf{R}_{ij}),
\label{eqBlake}
\end{eqnarray}
where $\mathbf{r}_{ij}= \mathbf{r}_i - \mathbf{r}_j$ is the vector between particles $i$ and $j$, and $\mathbf{R}_{ij}= \mathbf{r}_i - \overline{\mathbf{r}}_j$ is the vector between particle $i$ and the image of particle $j$ at $\overline{\mathbf{r}}_j = (x_j, y_j, -z_j)$. The Rotne-Prager tensor $\tensor{\mathbf{G}}^{\mathrm{RP}}$ is given as \cite{Rotne1969}
\begin{equation}
\tensor{\mathbf{G}}^{\mathrm{RP}} (\mathbf{r}) = \dfrac{1}{8\pi\eta |\mathbf{r}|} \left( \tensor{\mathbf{I}} + \dfrac{\mathbf{r} \otimes \mathbf{r}}{|\mathbf{r}|^2} \right) \nonumber + \dfrac{a^2}{4\pi\eta |\mathbf{r}|^3} \left( \dfrac{\tensor{\mathbf{I}}}{3} - \dfrac{\mathbf{r} \otimes \mathbf{r}}{|\mathbf{r}|^2} \right).
\label{eqRotnePrager}
\end{equation}
with fluid viscosity $\eta$. The last term in Eq. \ref{eqRotnePrager} involves the Rotne-Prager correction terms to the Stokes and source doublets, with its (off-)diagonal elements reading as \cite{hansen2011}
\begin{eqnarray}
\Delta G_{\alpha\alpha} &=& \dfrac{1}{4\pi\eta} \left[ \dfrac{-z_i z_j}{R^3} \left( 1-3\dfrac{R^2_{\alpha}}{R^2} \right) \right.  \\
 				      &+& \left. \dfrac{a^2R_z^2}{R^5} \left(1 - 5\dfrac{R^2_{\alpha}}{R^2} \right) \right], \\
\Delta G_{\alpha\beta} &=& \dfrac{1}{4\pi\eta} \left( \dfrac{3 z_i z_j R_{\alpha} R_{\beta}}{R^5} - 5a^2 \dfrac{R_{\alpha} R_{\beta} R^2_z}{R^7} \right),
\end{eqnarray}
where $\alpha, \beta \in \{x,y\}$, and $R_{\alpha}$, $R_{\beta}$ corresponding to $\alpha$- and $\beta$-component of $\mathbf{R}_{ij}$, and $z_i$ specifying the $z$-coordinate of particle $i$. Note that in simulations all particles have the same vertical distance $z$ from the substrate as adjusted according to the experimental gravitational length.

The no-slip boundary at the wall alters the particles' self-mobilities. We therefore employ a Rotne-Prager level self-mobility tensor $\tensor{\mathbf{G}}^{\mathrm{RPB}}_{self}(z) \equiv \mu^{\mathrm{RPB}}_{\|}(z) \tensor{\mathbf{I}}$ to obtain an expression for the dependance of the self-mobility of a colloid separated from a wall by distance $z$ with the diagonal element being \cite{hansen2011}
\begin{equation}
\mu^{\mathrm{RPB}}_{\|}(z) = \mu_0 \left( 1- \dfrac{9a}{16z} +  \dfrac{1}{8} \left( \dfrac{a}{z} \right)^3 \right) + O(a^4),
\end{equation}
and $\mu_0 = 1 / (6\pi\eta a)$ describing the Stokes self-mobility.

Finally, the equation for the trajectory $\textbf{r}_i$ of a colloidal particle $i$ obeying Brownian motion after a time step $\delta t$  reads 
\begin{equation}
\mathbf{r}_i (t+\delta t) = \mathbf{r}_i(t) + \left( \sum_{j=1}^{N} \tensor{\mathbf{\mu}}_{ij} \mathbf{F}_j \right) \delta t + \delta \textbf{W}_i,
\end{equation}
where $\tensor{\mathbf{\mu}}_{ij}$ is given as
\begin{equation}
\tensor{\mathbf{\mu}}_{ij} = \tensor{\mathbf{G}}^{\mathrm{RPB}}_{self}(z_i) \delta_{ij} + (1-\delta_{ij}) \tensor{\mathbf{G}}^{\mathrm{RPB}}(\mathbf{r}_i, \mathbf{r}_j), 
\end{equation}
which comprises both the self-mobility part and the entrainment of particle $i$ by the hydrodynamic flow-field created by conservative forces $\mathbf{F}_j$ acting on particle $j$. This force stems from the pair interactions, $V$, and for the $27$ driven wall particles also from the harmonic trap potential $V_t$. The random displacement $\delta \textbf{W}_i$ is sampled from a Gaussian distribution with zero mean and variance $2D_0 \delta t$ (for each Cartesian component) fixed by the fluctuation-dissipation relation, where $D_0$ is the diffusion coefficient.

In the simulations, the length scale is set by $\kappa$, the energy scale by $k_BT$, and the time scale by $\tau = 1/(\kappa^2 D_0)$. The inverse screening length $\kappa$ has been chosen as $\kappa a = 14.85$, where the experimental value of the radius served as a reference. Consequently, the corral radius has been set to $\kappa R_0=128$ yielding the experimental ratio of $R_0 / (2a) \approx 4.31$. The high screening at $\kappa a  = 14.85$ together with the contact potential chosen as $V(r=2a) \approx 0.85 k_BT$ ensures the quasi hard-disc-behaviour. Another crucial parameter in our system is the trap strength which has been set to $\lambda = 0.42 \kappa^2 k_BT$ in order to mimic the laser trap strength in the experiments. The time step is chosen as $\delta t = 10^{-4} \tau$. Our simulations run for up to $8 \times 10^3 \tau$, corresponding to approximately $35.5 \tau_B$ with $\tau_B$ being the experimental Brownian time, ie., the time one colloid needs to diffuse a length equal to its radius.

\section{Determining the Drag Coefficient}
\label{secDeterminingDrag}

\begin{figure*}[htb]
\begin{center}
\centerline{\includegraphics[width=140mm]{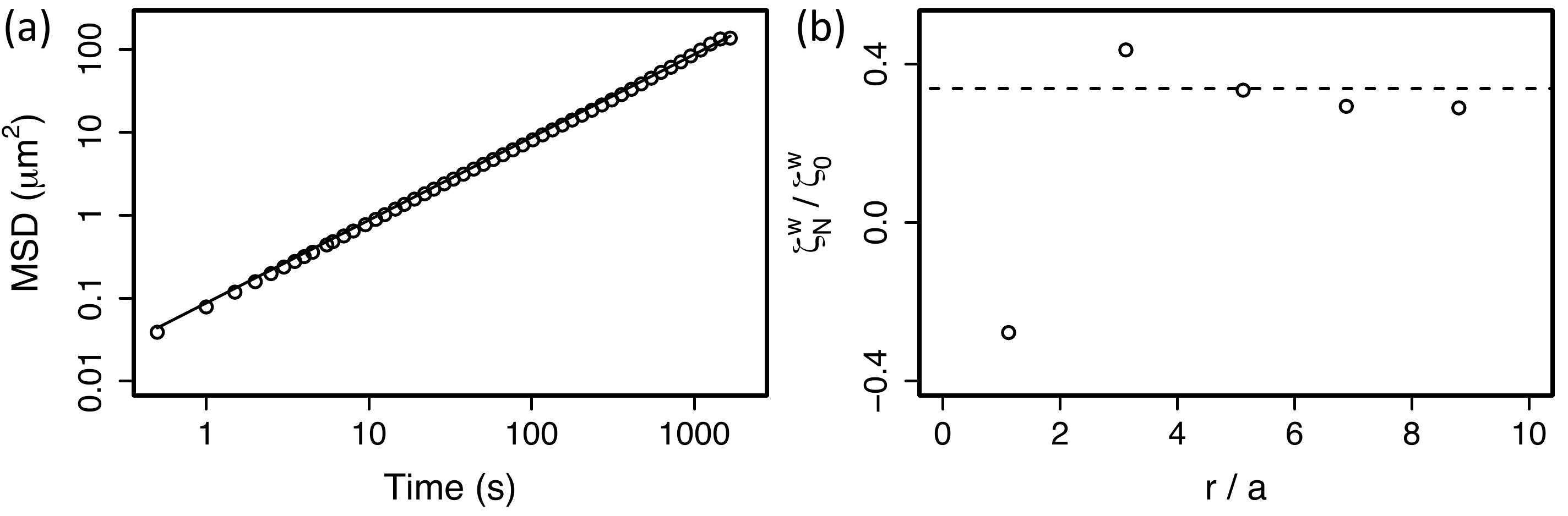}}
\caption{\label{figMSD} (a) Mean squared displacement in the dilute limit in the absence of optical traps. Line is linear fit used to extract the empirical drag coefficient. (b) Drag coefficient correction term of the form in \cite{ladavac2005} as a function of radial location, calculated using experimentally measured layered structure. Horizontal dashed line shows the average over layers 2 to 5.}
\end{center}
\end{figure*}

The analysis described in the next section relies on knowledge of the drag forces acting on each each particle to extract dimensional forces, and therefore stresses and viscosities. Thus, it is necessary to know the single particle drag coefficient for our $a = 2.5 \; \mathrm{\upmu m}$ radius polystyrene spheres undergoing quasi-two-dimensional motion near the substrate. To this end, we track the diffusive motion of these particles in the dilute limit, without any optical traps, and show the resulting mean squared displacement in Fig. \ref{figMSD} (a). The Stokes-Einstein relation says that this is linear in time, with gradient $4k_\mathrm{B}T/\zeta$. Thus, we perform a linear fit to the experimental data in Fig. \ref{figMSD} (a) and obtain an empirical measurement of the single particle drag coefficient to be $\zeta_\mathrm{emp} = 1.9 \times 10^{-7} \; \mathrm{kg \, s^{-1}}$. This represents an approximate four-fold increase over the Stokes drag coefficient $\zeta_0 = 6\pi \eta a = 4.7 \times 10^{-8} \; \mathrm{kg \, s^{-1}}$ for spheres of $2.5 \; \mathrm{\upmu m}$ radius in water.

However, it is known that multiple particles moving along a circular path of radius $r$ each experience reduced hydrodynamic drag due to the presence of the other particles \cite{ladavac2005}. This is the drafting effect. Ladavac \& Grier give an approximate result at the level of the Stokeslet approximation for the single-particle drag coefficient when $n$ particles of radius $a$ are equally spaced on a ring of radius $r$, where $r \gg a$, and near a planar substrate \cite{ladavac2005}.

\begin{widetext}
\begin{eqnarray}
\label{eqGrierDrag}	
\frac{\zeta_N^w}{\zeta_0^w} & = & \left\{1 + \frac{3 a \zeta_0^w}{8 r \zeta_0}
\sum_{j=2}^n \left[ \frac{(1 + \cos \theta_{1j})(1 - 3 \cos \theta_{1j})}{\sqrt{2 - 2 \cos \theta_{1j}}}\frac{h}{r} +  \frac{8 \cos \theta_{1j}}{(2 - 2 \cos \theta_{1j})^{3/2}} \frac{h^2}{r^2} + \frac{6(1+ \cos \theta_{1j})(5 \cos \theta_{1j} - 3)}{(2 - 2 \cos \theta_{1j})^{3/2}}\frac{h^3}{r^3} \right] \right\}^{-1}
\end{eqnarray}
\end{widetext}

\noindent
This result is truncated at order $(h/r)^3$. Here $\theta_{ij} = (2\pi/n)(j-i)$ is the angular separation between particles $i$ and $j$ and $h$ is the distance between the substrate and the particle centres.

Using the radial location of each layer's peak, $r_i$, and the layer populations $n_i$, we calculate this modification to the drag coefficient for our system, assuming $h = a + l_g$, that is that the particles are located one gravitational length above the substrate. The dependance on the radius of the circular path and the number of particles suggests that particles in different layers of our system will experience different drag coefficients. The results of this calculation are shown in Fig. \ref{figMSD} (b). The drag correction calculated using equation \ref{eqGrierDrag} for layer 1 is negative, an unphysical result that is a consequence of the fact that equation \ref{eqGrierDrag} is valid only for $r \gg a$, which is not true for layer 1 as $r_1 \approx a$. Therefore, ignoring the invalid layer 1 result, it is evident that the drafting effect results in a reduced drag coefficient compared to the dilute limit single-particle wall-corrected drag coefficient $\zeta_0^w$. Furthermore, the anticipated dependence of $\zeta_N^w$ on radial position is evident, though weak. If we identify our empirically measured wall-corrected drag coefficient, $\zeta_\mathrm{emp}$, with $\zeta_0^w$, then we anticipate that particles in our system will experience a drag coefficient of approximately $0.34 \, \zeta_\mathrm{emp}$ due to the drafting effect, where $0.34$ is the average of $\zeta_N^w / \zeta_0^w$ over layers 2 to 5, indicated by the dashed line in Fig. \ref{figMSD} (b).

Estimating the true drag correction is further complicated by the fact that our system consists of a series of concentric and closely interacting layers. Equation \ref{eqGrierDrag} considers a single circular particle layer in isolation, and therefore cannot be a true description of our system. The effective drag coefficient experienced by a particle in layer $i$ likely depends on the motion of particles in layers $i+1$ and $i-1$ in addition to the other particles in layer $i$. Therefore, the calculated reduction in drag is really only an order-of-magnitude estimate of the drafting effect. The sum in equation \ref{eqGrierDrag} is dominated by contributions from particles $j=2$ and $j=n$, which are the neighbouring particles of particle $1$. Since the separation between neighbouring particles is approximately the same in all layers, and since the radial dependence of of $\zeta_N^w$ is predicted to be weak, we treat the drag coefficient as being independent of radial position and equal to $\zeta = 0.34 \, \zeta_\mathrm{emp}$. This assumption is necessary to estimate the drag coefficient in layer 1, for which equation \ref{eqGrierDrag} is invalid, as indicated its prediction of a negative drag coefficient in this region.

\section{Interlayer Hopping at Lower Population}
\label{secInterlayer}

\begin{figure*}[htb]
\begin{center}
\centerline{\includegraphics[width=140mm]{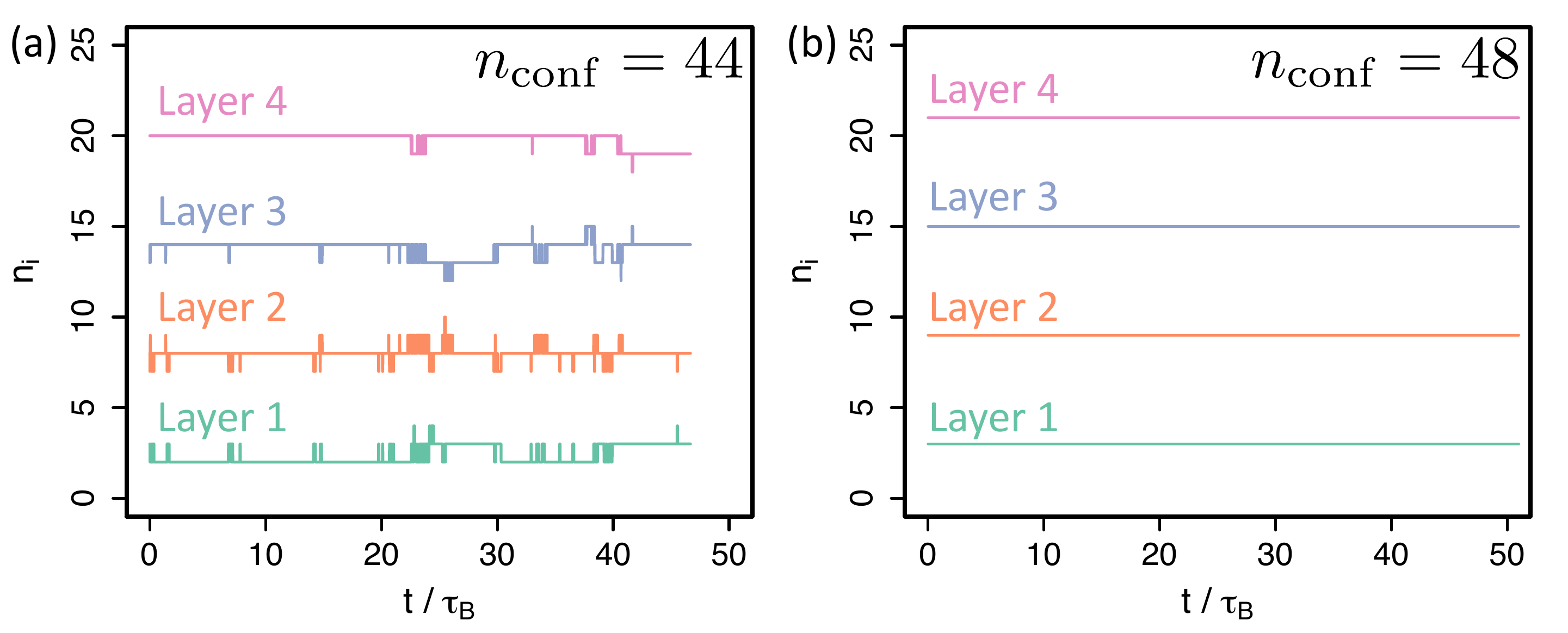}}
\caption{\label{figParticleCountLayers} Time evolution of layer populations $n_i$ measured in experiments driven at $\Pecl = 17.5$ with confined populations (a) $n_\mathrm{conf} = 44$ and (b) $n_\mathrm{conf}=48$.}
\end{center}
\end{figure*}

Figure \ref{figParticleCountLayers} shows the time evolution of layer populations in experiments driven at $\Pecl = 17.5$ for confined populations (a) $n_\mathrm{conf} = 44$ and (b) $n_\mathrm{conf}=48$. In the more densely packed system at $n_\mathrm{conf}=48$ the populations of all layers are constant in time, as required for the rheological analysis described in the main manuscript. When the population is reduced to $n_\mathrm{conf} = 44$, however, particles occasionally move between layers. When layer populations are not fixed, the rheological analysis as described in the main manuscript cannot be applied, and so we have focused our attention on the $n_\mathrm{conf} = 48$ system.

\end{document}